\documentclass[letter,bibyear]{aa} 
\usepackage{epsfig}
\usepackage{amsmath}
\usepackage{subfigure}
\usepackage{natbib}
\usepackage{multirow}
\usepackage{color}
\usepackage{longtable}
\usepackage{graphicx}
\usepackage{epstopdf}
\usepackage{booktabs}
\usepackage{txfonts}

\newcommand{\jmst}{J.~Mol.~Struct.}   
\newcommand{\chemrev}{Chem.~Rev.}

\begin{document}

\title{Space and laboratory discovery of HC$_3$S$^+$
\thanks{Based on observations carried out with the Yebes 40m telescope (projects 19A003, 20A014, and 20D015). 
The 40m radiotelescope at Yebes Observatory is operated by the Spanish Geographic Institute (IGN, Ministerio de 
Transportes, Movilidad y Agenda Urbana).}
}

\author{
J.~Cernicharo\inst{1},
C.~Cabezas\inst{1},
Y.~Endo\inst{2},
N.~Marcelino\inst{1},
M.~Ag\'undez\inst{1},
B.~Tercero\inst{3,4}, 
J.~D.~Gallego\inst{4}, and
P.~de~Vicente\inst{4}
}

\institute{Grupo de Astrof\'isica Molecular, Instituto de F\'isica Fundamental (IFF-CSIC), C/ Serrano 121, 28006 Madrid, Spain
\email jose.cernicharo@csic.es
\and Department of Applied Chemistry, Science Building II, National Chiao Tung University, 1001 Ta-Hsueh Rd., Hsinchu 30010, Taiwan
\and Observatorio Astron\'omico Nacional (IGN), C/ Alfonso XII, 3, 28014, Madrid, Spain
\and Centro de Desarrollos Tecnol\'ogicos, Observatorio de Yebes (IGN), 19141 Yebes, Guadalajara, Spain
}

\date{Received; accepted}

\abstract{We report the detection in TMC-1 of the protonated form of C$_3$S. The discovery of the cation 
HC$_3$S$^+$ was carried through the observation of four harmonically related lines in the Q band using the 
Yebes 40m radiotelescope, and is supported by accurate \textit{ab initio} calculations and laboratory 
measurements of its rotational spectrum. We derive a column density  $N$(HC$_3$S$^+$) = 
$(2.0\pm0.5)\times10^{11}$ cm$^{-2}$, which translates to an abundance ratio  C$_3$S/HC$_3$S$^+$ 
of 65 $\pm$ 20. This ratio is comparable to the CS/HCS$^+$ ratio (35 $\pm$ 8) and is a factor of 
about ten larger than the C$_3$O/HC$_3$O$^+$ ratio previously found in the same source. However, the 
abundance ratio HC$_3$O$^+$/HC$_3$S$^+$ is 1.0 $\pm$ 0.5, while C$_3$O/C$_3$S is just $\sim0.11$. 
We also searched for protonated C$_2$S in TMC-1, based on \textit{ab initio} calculations of its 
spectroscopic parameters, and derive a 3$\sigma$ upper limit of $N$(HC$_2$S$^+$) $\leq$ 
$9\times10^{11}$ cm$^{-2}$ and a C$_2$S/HC$_2$S$^+$ $\geq$ 60. The observational results are 
compared with a state-of-the-art gas-phase chemical model and conclude that HC$_3$S$^+$ is mostly 
formed through several pathways: proton transfer to C$_3$S, reaction of S$^+$ with $c$-C$_3$H$_2$, 
and reaction between neutral atomic sulfur and the ion C$_3$H$_3^+$.}


\titlerunning{HC$_3$S$^+$ in TMC-1}
\authorrunning{Cernicharo et al.}

\maketitle

\section{Introduction}

The cold dark core TMC-1 presents an interesting carbon-rich chemistry that leads to the formation of 
long neutral carbon-chain radicals and their anions (see \citealt{Cernicharo2020a} and references therein). 
The carbon chains C$_2$S and C$_3$S are particularly abundant in this cloud \citep{Saito1987,Yamamoto1987}, 
as they also are in the carbon-rich circumstellar envelope IRC\,+10216 \citep{Cernicharo1987}. TMC-1 is 
also peculiar for the presence of protonated species of abundant carbon chains, such as HC$_3$NH$^+$
\citep{Kawaguchi1994}, HC$_3$O$^+$ 
\citep{Cernicharo2020a}, and HC$_5$NH$^+$ \citep{Marcelino2020}. The abundance ratio between a protonated 
molecule and its neutral counterpart, [MH$^+$]/[M], is sensitive to the degree of ionisation, and therefore also to 
various physical parameters of the cloud,
as well as to the formation and destruction rates of the cation \citep{Agundez2015}. It is interesting to 
note that both chemical models and observations suggest a trend in which the abundance ratio [MH$^+$]/[M] 
increases with increasing proton affinity of M \citep{Agundez2015}. Thus, protonated species of abundant 
molecules with high proton affinities are good candidates for detection. This is the case for HC$_3$S$^+$, 
the sulphur analogue of HC$_3$O$^+$, given that C$_3$S has a very high proton affinity of 933 kJ mol$^{-1}$
\citep{Hunter1998}, and is around seven times more abundant than C$_3$O.

In this letter, we report the detection of four harmonically related lines that belong to a molecule 
with a $^1\Sigma$ ground electronic state towards the cold dark core TMC-1. 
From the derived rotational and distortion constants we conclude, based on detailed
\emph{ab initio} calculations,
that the best possible carrier is HC$_3$S$^+$, the protonated form of C$_3$S. We succeeded in
producing this cation in the laboratory and measured its microwave spectrum, which fully confirms our 
assignment.
Previous laboratory studies of this species were performed by \citet{Thorwirth2020}
who measured its vibrational spectrum through infrared observations at low spectral resolution.
We present a detailed observational study of protonated S-bearing carbon chains in this 
cloud and discuss these results in the context of a state-of-the-art gas-phase chemical model.

\section{Observations}

New receivers built within the Nanocosmos project\footnote{\texttt{https://nanocosmos.iff.csic.es/}} 
and installed at the Yebes 40m radiotelescope were used
for observations of TMC-1. The Q-band receiver consists of two HEMT cold amplifiers covering the 
31.0-50.3 GHz band with horizontal and vertical polarizations. Receiver temperatures vary from 22 K at 32 GHz 
to 42 K at 50 GHz. The spectrometers are $2\times8\times2.5$ GHz FFTs with a spectral resolution of 38.15 kHz 
providing the whole coverage of the Q-band in both polarisations. The main beam efficiency varies from 0.6 at 
32 GHz to 0.43 at 50 GHz. A detailed description of the system is given by \citet{Tercero2021}.

The line survey of TMC-1 ($\alpha_{J2000}=4^{\rm h} 41^{\rm  m} 41.9^{\rm s}$ and $\delta_{J2000}=+25^\circ 41' 27.0''$) 
in the Q-band was performed in several sessions. Previous results on the detection of C$_3$N$^-$ and C$_5$N$^-$ 
\citep{Cernicharo2020b}, HC$_5$NH$^+$ \citep{Marcelino2020}, HC$_4$NC \citep{Cernicharo2020c}, and HC$_3$O$^+$ 
\citep{Cernicharo2020a} were based on two runs performed in November 2019 and February 2020. In these runs, two 
different frequency coverages were observed, 31.08-49.52 GHz and 31.98-50.42 GHz, which allow us to check that no 
spurious ghosts are produced in the down-conversion chain in which the signal coming from the receiver is 
down-converted to 1-19.5 GHz, and then split into eight bands of 2.5 GHz, each of which are analyzed 
by the FFTs. Additional data were taken in October and December 2020
to improve the line survey at some frequencies, and to 
further check the consistency of all observed spectral features.

The observing procedure was frequency-switching with a frequency throw of 10\,MHz for the two first runs and of 
8\,MHz for those of October and December 2020. The intensity scale, antenna temperature ($T_A^*$), was calibrated using two absorbers 
at different temperatures and the atmospheric transmission model ATM \citep{Cernicharo1985, Pardo2001}. Calibration 
uncertainties have been adopted to be 10~\%. The nominal spectral resolution of 38.15 kHz was used for the final 
spectra. The sensitivity varies along the Q-band between 0.5 and 2.5 mK, which considerably improves previous line 
surveys in the 31-50 GHz frequency range \citep{Kaifu2004}. All data were analysed using the GILDAS 
package\footnote{\texttt{http://www.iram.fr/IRAMFR/GILDAS}}.

\section{Results} \label{results}

\begin{figure}
\centering
\includegraphics[width=0.95\columnwidth,angle=0]{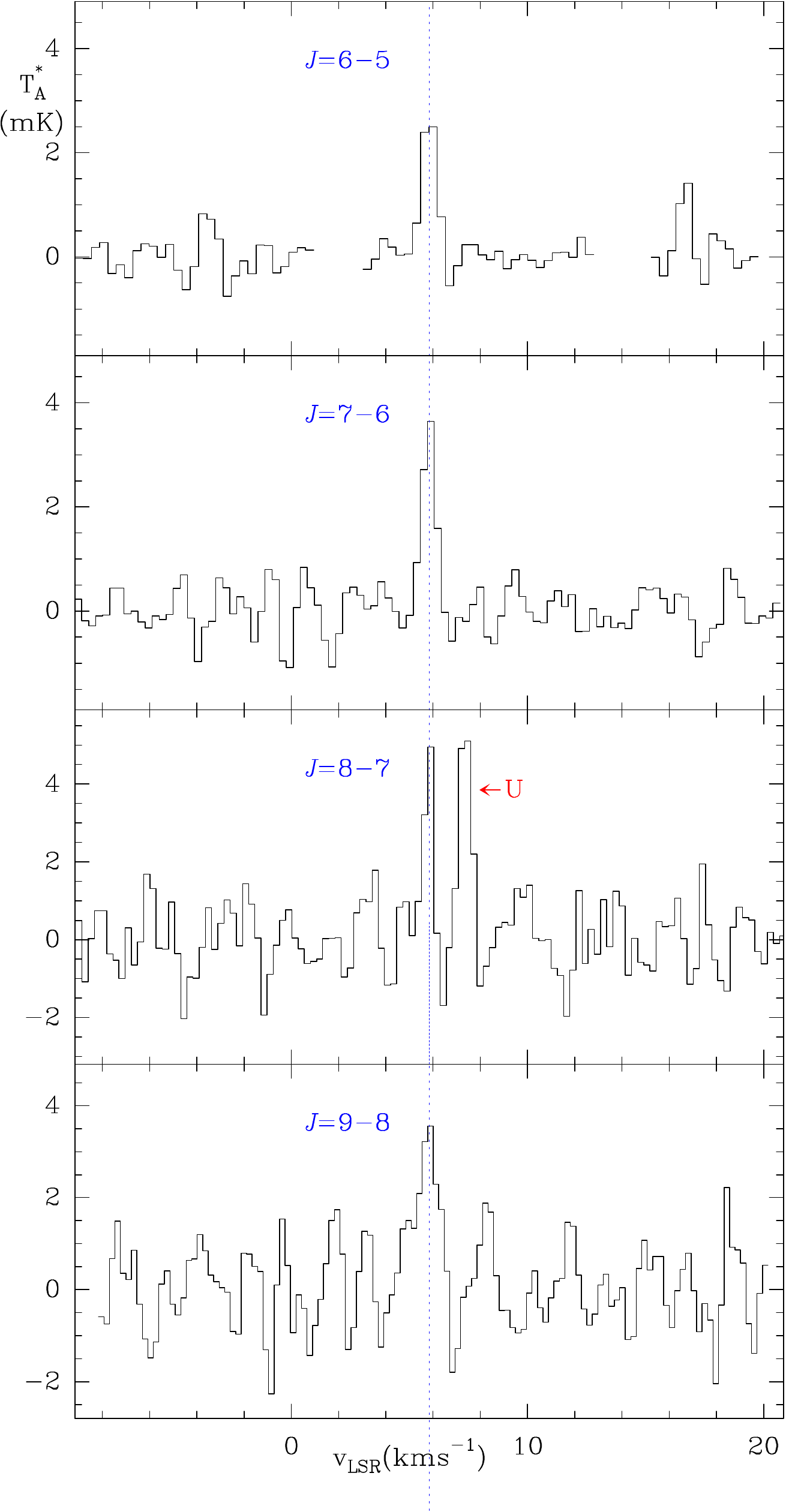} 
\caption{
Observed lines of the new molecule found in the 31-50 GHz domain towards TMC-1. The abscissa corresponds 
to the local standard of rest velocity
in km s$^{-1}$. Frequencies and intensities for the observed lines are given in Table \ref{tab_hc3s+}. The ordinate 
is the antenna temperature corrected for atmospheric and telescope losses in mK. Spectral resolution is 38.15 kHz. 
Blanked channels in the top panel correspond to negative features produced in the folding of the frequency-switching 
observations.}
\label{fig_hc3s+}
\end{figure}

\begin{table}
\tiny
\caption{Observed line parameters for HC$_3$S$^+$ in the laboratory and in TMC-1}
\label{tab_hc3s+}
\centering
\begin{tabular}{cccccc}
\hline
{\textit J$_u$-$J_l$}& $\nu_{obs}$~$^a$& $\nu_{o}$$-$$\nu_{c}$~$^b$  & $\int T_A^* dv$~$^c$ & $\Delta v$~$^d$ & $T_A^*$ \\
                     &  (MHz)       &     (kHz)            & (mK\,km\,s$^{-1}$)      & (km\,s$^{-1}$)  & (mK) \\
\hline
3-2$^e$&  16412.7583& -1.8 & ... & ... & ...\\
4-3$^e$&  21883.6632&  2.4 & ... & ... & ...\\
5-4$^e$&  27354.5443& -0.8 & ... & ... & ...\\
6-5    &  32825.4090&  0.0 & 1.71$\pm$0.40 & 0.61$\pm$0.12& 2.6$\pm$0.4 \\
7-6    &  38296.2400& -7.5 & 1.74$\pm$0.30 & 0.52$\pm$0.09& 3.1$\pm$0.4 \\
8-7    &  43767.0670&  9.5 & 2.05$\pm$0.30 & 0.40$\pm$0.15& 3.9$\pm$0.4 \\
9-8    &  49237.8310& -3.4 & 3.20$\pm$0.95 & 0.79$\pm$0.15& 3.6$\pm$0.9 \\
\hline
\end{tabular}
\tablefoot{ 
\tablefoottext{a}{Observed frequencies in the laboratory (below 30 GHz) and towards TMC-1 (above 30 GHz). For this source 
we adopted a v$_{\rm LSR}$ of 5.83 km s$^{-1}$ \citep{Cernicharo2020a,Cernicharo2020b,Cernicharo2020c}. The 
frequency uncertainty is 3 and 10 kHz for the laboratory and astronomical lines, respectively.}\\
\tablefoottext{b}{Observed minus calculated frequencies in kHz. The calculated frequencies are obtained
 from the merged fit to the astronomical and laboratory measurements (see Sect. 5).} \\
\tablefoottext{c}{Integrated line intensity in mK\,km\,s$^{-1}$.} \\
\tablefoottext{d}{Line width at half intensity derived by fitting a Gaussian function to 
the observed line profile (in km\,s$^{-1}$).} \\
\tablefoottext{e}{Line observed in the laboratory (see Sect. \ref{laboratory}).}}\\
\end{table}
\normalsize

Line identification in our survey of TMC-1 was performed using the MADEX catalogue \citep{Cernicharo2012}, 
the Cologne Database of Molecular Spectroscopy catalogue (CDMS; \citealt{Muller2005}), and the JPL catalogue \citep{Pickett1998}. 
Among the unidentified lines in our survey we found four lines in nearly perfect harmonic relation 6:7:8:9. 
The lines are 
shown in Fig. \ref{fig_hc3s+} and the derived line parameters are given in Table \ref{tab_hc3s+}. The rotational and distortion 
constants derived from a fit to these lines are $B$ = 2735.4630 $\pm$ 0.0012 MHz and $D$ = 0.171 $\pm$ 0.009 kHz
(see column Exp. (TMC-1) in Table~\ref{abini}). The rotational 
constant is slightly below that of C$_3$S ($B$ = 2890.4 MHz) and slightly higher than that of HC$_3$S ($B$ = 2688.4 MHz; 
\citealt{Hirahara1994}) and of HC$_3$P ($B$ = 2656.4 MHz; \citealt{Bizzocchi2001}). In our previous discovery of HC$_3$O$^+$ 
in this source, we performed calculations for several protonated species of abundant neutral molecules in order to search for 
them in our survey. HC$_3$S$^+$ was one of the best candidates for detection in TMC-1. \textit{Ab initio} calculations by 
\citet{Thorwirth2020} indicate a rotational constant for this species of 2734.5 MHz, which is very close to the value we have 
derived for the new molecule. We made additional calculations at a different level of theory (see Section \ref{abinitio}) 
and found that the best prediction for $B$ and $D$ of HC$_3$S$^+$ (see Table~\ref{abini}) match those of the 
new species perfectly. Laboratory measurements (see Section \ref{laboratory}) confirm that the four observed lines belong 
to HC$_3$S$^+$.

From the line parameters in Table \ref{tab_hc3s+}, adopting a dipole moment of 1.73 D (see Sect.~\ref{abinitio}), and assuming 
a uniform source with a radius of 40$''$ \citep{Fosse2001}, we derive a column density of $N$(HC$_3$S$^+$) = 
$(2.0\pm0.4)\times10^{11}$ cm$^{-2}$ and a rotational temperature of $10\pm2$ K. This value is compatible with 
the upper limit of $3\times10^{11}$ cm$^{-2}$ obtained by \cite{Cernicharo2020a} from a search based on \textit{ab initio} 
calculations. The sensitivity improvement added by the new data has permitted its detection in TMC-1.

Protonated C$_2$S was predicted to be abundant in cold dense clouds by \cite{Agundez2015} based on its high proton affinity 
and large abundance of C$_2$S. We therefore carried out \textit{ab initio} calculations for HC$_2$S$^+$, which has a 
$^3\Sigma$ ground electronic state (see also \citealt{Puzzarini2008}), to search for the lines $N_N \rightarrow N-1_{N-1}$ ($J=N$), 
which could be in good harmonic relation. Our estimates for $B$ and $D$ are 6048 MHz and 0.12 kHz, respectively. 
These lines should exhibit a weak hyperfine splitting of $\sim$0.8 MHz. Two of 
these lines (3$_3$-2$_2$ and 4$_4$-3$_3$) fall within our Q-band survey. However, we have not found two harmonically
related features (3:4) that could
be attributed to them. The explored ranges for the 3$_3$-2$_2$ and 4$_4$-3$_3$ transitions are $\pm$40 MHz around
36288 and 48384 MHz, respectively. The sensitivity of our data in these ranges is $\sim$0.5 and $\sim$1 mK, respectively.
The other components of each triplet ($J=N\pm1$) will show a more complex pattern due to the spin--spin and
spin--rotation 
interaction. We adopted the $\lambda$ parameter from C$_2$S, which is isoelectronic with HC$_2$S$^+$, but the predictions 
have a large uncertainty and therefore a radioastronomical search is not straightforward for these transitions. 
Using the dipole moment 
derived in our \textit{ab initio} calculations (2.67 D), and assuming a rotational temperature similar
to that of CCS (5 K), we derive a 3$\sigma$ upper limit for its column density $\leq$9$\times10^{11}$ cm$^{-2}$. For T$_r$=10 K, the
3$\sigma$ upper limit is 3$\times$10$^{11}$ cm$^{-2}$. Nevertheless, the  lines searched here are not best suited for a 
detection of HC$_2$S$^+$
as they are expected to be much higher in energy that those of the $J=N+1$ series, which, as mentioned above,
require a good estimate of the spin--spin interaction constant, $\lambda$, to obtain reliable frequencies with which to carry out a search for this molecule.

We also derived column densities for S-bearing molecules related to HC$_3$S$^+$ using our Q-band line survey of TMC-1. 
The line parameters are given in Table~\ref{tab_s_bearing}. For CS and HCS$^+$ we only observed the $J$ = 1-0 transition 
and therefore we adopted a rotational temperature of 10 K. A similar approach was taken by \citet{Vastel2018} in their 
study of L1544. The column density of CS, whose $J$=1-0 line has a significant optical depth, was derived from that 
of C$^{34}$S adopting the $^{32}$S/$^{34}$S abundance ratio of 25$\pm$5 determined from C$_3$S and C$_3$$^{34}$S. The derived 
column densities for all species studied in this paper are given in Table \ref{column_densities}. 
The derived column densities for CCS and C$_3$S are in good agreement with those derived by \citet{Saito1987}
and \citet{Yamamoto1987}.
A detailed analysis
of the effect of the assumed rotational temperature on diatomic or linear polyatomic molecules is provided in Appendix \ref{LTE}.

\section{Quantum chemical calculations for HC$_3$S$^+$} \label{abinitio}

\begin{table*}
\small
\caption{Theoretical and experimental values for the spectroscopic parameters of HC$_3$S$^+$ (all in MHz).}
\label{abini}
\centering
\begin{tabular}{{lccccccc}}
\hline
\hline
&\multicolumn{2}{c}{HC$_3$P}&\multicolumn{5}{c}{HC$_3$S$^+$} \\
\cmidrule(lr){2-3} \cmidrule(lr){4-8}
Parameter & Calc.\tablefootmark{a} & Exp.\tablefootmark{b} & Calc.\tablefootmark{a} & Scaled\tablefootmark{c} & Exp. (TMC-1)\tablefootmark{d}&Exp. (Labo)\tablefootmark{d}&Exp. (Merged)\tablefootmark{d} \\
\hline
$B_e$             &  2658.238 &                        &   2736.349        &               &               &               &                   \\
Vib-Rot. Corr.    &   2.430   &                        &   1.801           &               &               &               &                   \\
$B_0$             & 2655.808  &      2656.393295(52)   &   2734.548        &     2735.150  &  2735.4630(12)& 2735.46310(70)&  2735.46311(23)   \\
$D$ $\times 10^3$   & 0.171     &       0.1810132(39)    &    0.162        &     0.171     &    0.171(9)   &   0.172(17)   &    0.1720(29)      \\
\hline
\end{tabular}
\tablefoot{Numbers in parentheses are 1$\sigma$ uncertainties in units of the last digits. \\
\tablefoottext{a}{$B$ and $D$ derived from \emph{ab initio calculations} in this work.}\\ 
\tablefoottext{b}{\citet{Bizzocchi2001}.} \\
\tablefoottext{c}{This work; scaled by the ratio Exp/Calc. of the corresponding parameter for HC$_3$P species}. \\
\tablefoottext{d}{Experimental rotational and distortion constants derived in this work.}\\
}
\end{table*}

In order to obtain precise geometries and spectroscopic molecular parameters that help in the assignment of 
the observed lines we carried out high-level \emph{ab initio} calculations for HC$_3$S$^+$ using the 
Molpro 2018.1 \citep{Werner2018} and Gaussian09 \citep{Frisch2013} program packages. We followed the same 
strategy used previously for HC$_3$O$^+$ \citep{Cernicharo2020a}, whereby we scaled our calculations with 
a molecular system that is isoelectronic to the target molecule. In the present case, we chose HC$_3$P (see 
also Thorwirth et al. 2020), 
whose rotational 
parameters have been experimentally determined by \citet{Bizzocchi2001}, as a reference system to scale the 
HC$_3$S$^+$ calculations. The geometry optimisation calculations were carried out at CCSD(T)-F12/cc-pCVTZ-F12 level of theory 
\citep{Raghavachari1989,Adler2007,Knizia2009,Hill2010a,Hill2010b}, which has proven to be a suitable method with which
to accurately reproduce the molecular geometry of analogue molecules \citep{Cernicharo2019,Cernicharo2020a}. 
We first calculated $B_e$ for HC$_3$P and then computed $B_0$ using the zero-point vibrational contribution
calculated at MP2/cc-pVTZ level of theory. The agreement with the experimental value is very good, with a relative 
error of 0.02~\% (see Table~\ref{abini}). The $B_0$ value for HC$_3$S$^+$ was calculated using the $B_e$ value 
and the zero-point vibrational correction, obtained at CCSD(T)-F12/cc-pCVTZ-F12 and MP2/cc-pVTZ levels of theory, 
respectively. This value was then corrected using a scaling factor 
obtained from the ratio between the experimental and theoretical values derived for HC$_3$P. 
The final value of $B_0$ obtained for HC$_3$S$^+$ agrees very well with that obtained from observations 
and in the laboratory, with a relative error of around 0.01~\%. In addition, the centrifugal distortion value obtained using the same procedure 
at MP2/cc-pVTZ level of theory is compatible with that obtained from the fit of the lines. The computed dipole moment 
is 1.73\,D. The results of our calculations are in agreement with those obtained by \citet{Thorwirth2020}, which were 
made using the same procedure but at a different level of theory.

\begin{figure}
\includegraphics[scale=0.40]{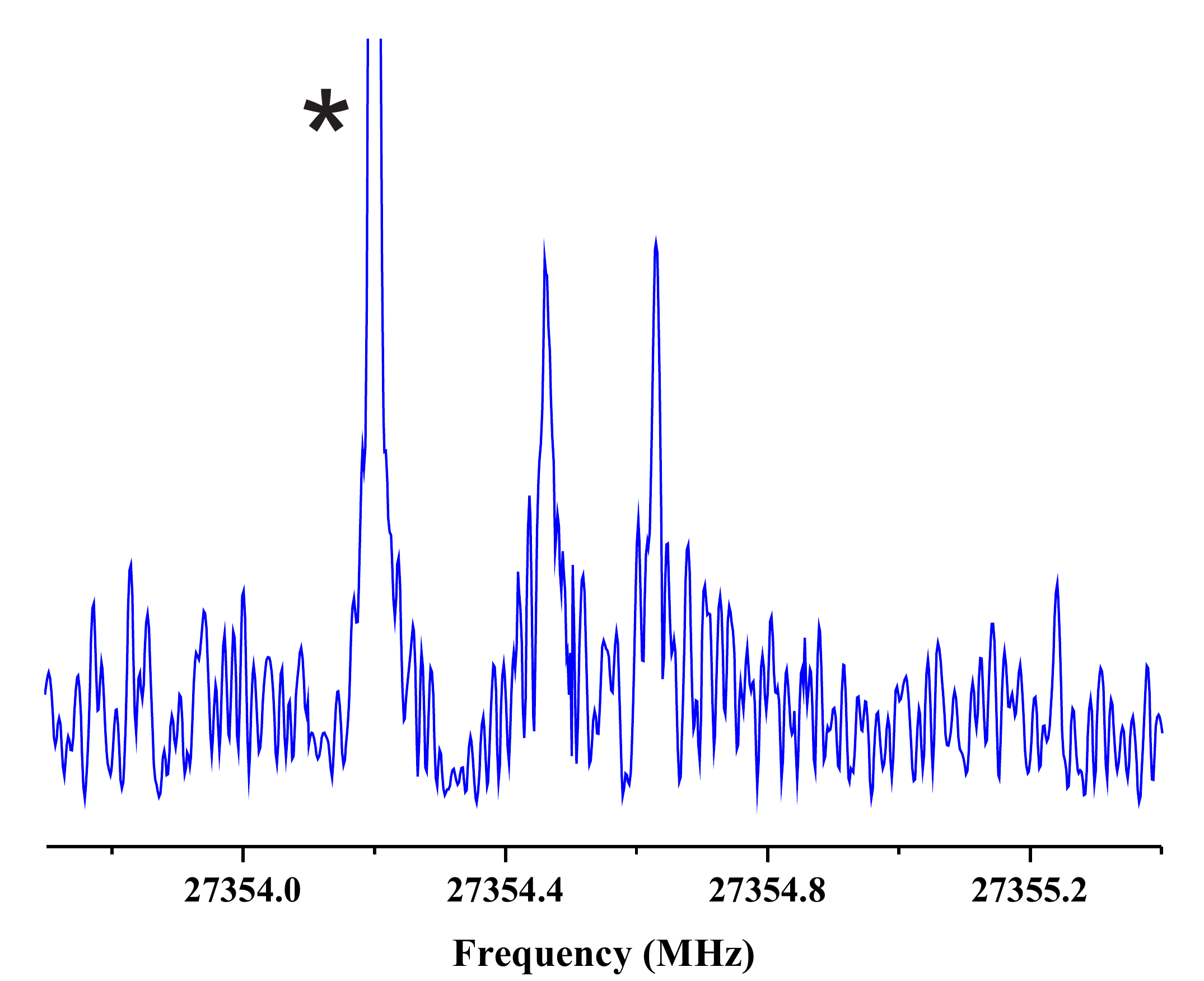}
\centering
\caption{\label{FTMW_spe} FTMW spectra of HC$_3$S$^+$ showing the $J$ = 5-4 rotational transition at 27.3 GHz. 
The spectrum was achieved by 10000 shots of accumulation at a repetition rate of 10 Hz. The coaxial arrangement of the 
adiabatic expansion and the resonator axis produces an instrumental Doppler doubling. The resonance frequency is calculated 
as the average of the two Doppler components. The feature marked with an asterisk is an artifact.}
\end{figure}

\section{Laboratory detection of HC$_3$S$^+$} \label{laboratory}

The rotational spectrum of HC$_3$S$^+$ was measured using a Balle-Flygare-type Fourier transform microwave (FTMW) 
spectrometer combined with a pulsed discharge nozzle \citep{Endo1997,Cabezas2016}, which was previously used 
to characterise other highly reactive molecules. The transient species, HC$_3$S$^+$, was produced in a 
supersonic expansion by a pulsed electric discharge of a gas mixture 
of C$_2$H$_2$ (0.2\%), H$_2$ (5\%), and CS$_2$ (0.2\%) diluted 
in Ne and applying a voltage of 1100 V through the throat of the nozzle source. The rotational constants derived from 
the astronomical observations were used to predict the frequencies of the rotational transitions $J$ = 3-2, 4-3, and 
5-4 of HC$_3$S$^+$. A scan of $\pm$2 MHz was achieved around these frequencies and three lines were observed 
at 16412.7583, 21883.6632, and 27354.5443 MHz with un uncertainty of 3 kHz (see Fig.~\ref{FTMW_spe} and Table~\ref{tab_hc3s+}), 
just 1-5 kHz away from the predicted frequencies using the derived rotational and distortion
constants from the TMC-1 observations.
The following experimental results confirm that these lines belong to a transient species: 
(i) they disappear in the absence of electric discharge, and (ii) the lines disappear when CS$_2$ is removed from the gas mixture. 
No more lines at lower or higher frequencies ($J$ = 2-1 and 6-5) could be observed due to the spectrum weakness and the 
poorer performance of the spectrometer at those frequencies.

The fitted rotational and distortion constants derived from the 
laboratory data alone are
given in column $Exp.$ $(labo)$ of Table \ref{abini}.
A merged fit  to the laboratory and astronomical frequencies provides a rotational constant 
$B$=2735.46311$\pm$0.00023 MHz and a distortion constant $D$=0.1720$\pm$0.0029 kHz. The 
standard deviation of the fit is 5.8 kHz and the correlation coefficient between $B$ and $D$ is 0.82. 
These are the recommended constants to predict the rotational spectrum of HC$_3$S$^+$. 
The predicted frequencies, Einstein coefficients, upper energy levels, and line
strengths for rotational transitions with $J\leq30$ are given in Table \ref{pred_hc3s+}.
The observed 
minus calculated frequencies from this merged fit are given in Table \ref{tab_hc3s+}.

\begin{table}
\caption{Column densities derived for S-bearing species in TMC-1}
\label{column_densities}
\centering
\begin{tabular}{{lcc}}
\hline
Molecule         &               $N$ (cm$^{-2}$)                   & $T_{rot}$ (K) \\
\hline
CS~$^a$          & (3.50$\pm$0.40)$\times$10$^{14}$  & 10.0~$^b$ \\
C$^{34}$S        & (1.45$\pm$0.10)$\times$10$^{13}$  & 10.0~$^b$ \\
$^{13}$C$^{34}$S & (1.45$\pm$0.20)$\times$10$^{11}$  & 10.0~$^b$ \\
HCS$^+$          & (1.00$\pm$0.10)$\times$10$^{13}$  & 10.0~$^b$ \\
HC$^{34}$S$^+$   & (7.10$\pm$0.70)$\times$10$^{11}$  & 10.0~$^b$ \\
CCS              & (5.50$\pm$0.65)$\times$10$^{13}$  &  5.1$\pm$0.2\\
CC$^{34}$S       & (5.00$\pm$0.50)$\times$10$^{12}$  &  3.6$\pm$0.4\\
CCCS             & (1.30$\pm$0.13)$\times$10$^{13}$  &  5.8$\pm$0.2\\
CCC$^{34}$S      & (5.30$\pm$0.50)$\times$10$^{11}$  &  6.7$\pm$0.2\\
HC$_3$S$^+$      & (2.00$\pm$0.40)$\times$10$^{11}$  & 10.0$\pm$2.0\\
HCCS$^+$~$^c$         & $\leq$9.00$\times$10$^{11}$       & 10.0~$^b$ \\
\hline
\end{tabular}
\tablefoot{\tablefoottext{a}{Derived from C$^{34}$S and the C$_3$S/C$_3$$^{34}$S abundance ratio.}\\ 
\tablefoottext{b}{Rotational temperature has been fixed. See Appendix A.}\\
\tablefoottext{c}{3$\sigma$ upper limit.}
}
\end{table}

\section{Chemical model}

\begin{figure*}
\centering
\includegraphics[width=\textwidth,angle=0]{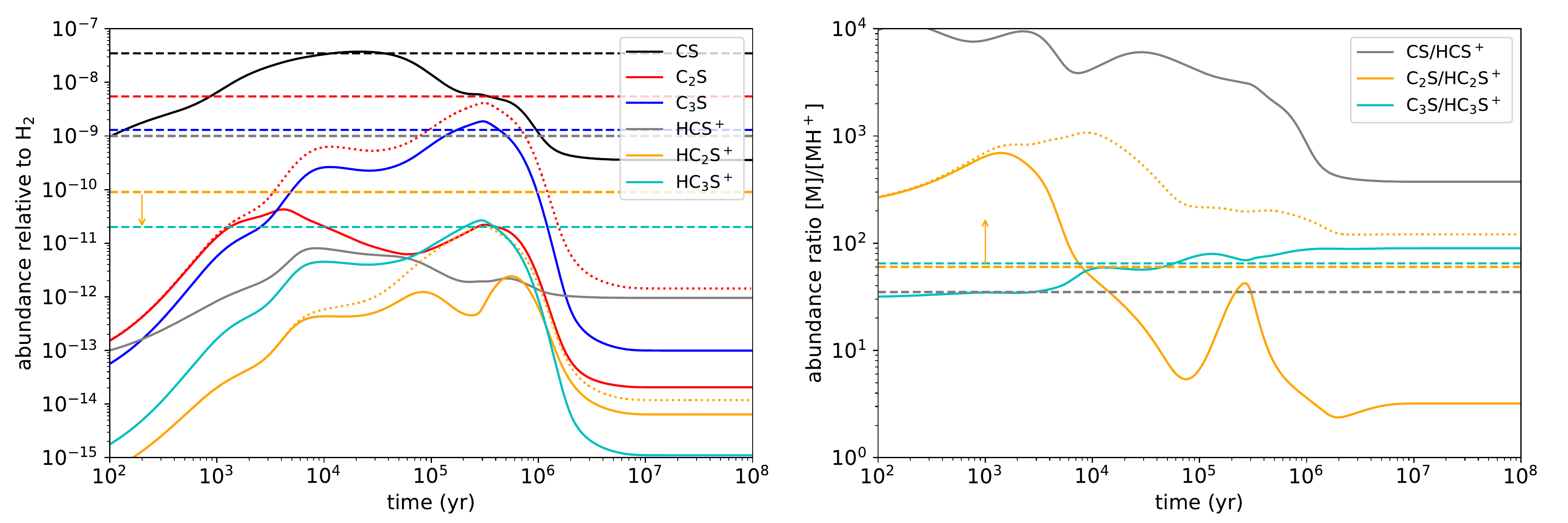}
\caption{Calculated abundances (left) and abundance ratios (right) as a function of time 
for a cold dense cloud. Observed values in TMC-1 (see Table~\ref{column_densities}) are indicated by 
horizontal dashed lines. The observed HC$_2$S$^+$ abundance and C$_2$S/HC$_2$S$^+$ ratio are an upper 
and a lower limit, respectively, and are indicated by vertical arrows. Dotted lines represent the 
calculated abundances of C$_2$S and HC$_2$S$^+$ (left panel) and its ratio (right panel) when the 
reaction O  + C$_2$S is neglected.}
\label{fig:model}
\end{figure*}

From Table~\ref{column_densities} we derive the following neutral-to-protonated column density ratios: 
CS/HCS$^+$ = 35 $\pm$ 6, C$_2$S/HC$_2$S$^+$ $\geq$ 60, 
and C$_3$S/HC$_3$S$^+$ = 65 $\pm$ 20. 
In order 
to interpret these ratios we carried out chemical model calculations similar to those presented by 
\cite{Agundez2015}, who discussed the chemistry of protonated molecules in cold dense clouds. We have 
adopted the KIDA {\small \texttt{kida.uva.2014}} chemical network \citep{Wakelam2015}. The abundances of 
sulfur-bearing molecules are strongly dependent on the depletion of sulfur, which is still a matter of 
debate \citep{Vidal2017,Vastel2018}. Here we have adopted the so-called set of low-metal elemental abundances 
(see \citealt{Agundez2013}), in which S/H = $8\times10^{-8}$. The resulting fractional abundances and abundance 
ratios are shown in Fig.~\ref{fig:model} as a function of time. If we focus on the so-called early time 
(10$^5$-10$^6$ yr), at which calculated abundances agree better with observations (e.g. \citealt{Agundez2013}), 
we see that the chemical model reproduces the observed fractional abundances of neutral and
protonated molecules  reasonably well, at the exception of C$_2$S and HCS$^+$, which translates to good agreement between calculated 
and observed [M]/[MH$^+$] ratios for M = C$_3$S but not for M = CS and C$_2$S.

As discussed by \cite{Agundez2015}, in a simplified chemical scheme, a protonated molecule is formed through proton 
transfer from a proton donor (typically HCO$^+$, H$_3$O$^+$, and H$_3^+$) and destroyed by dissociative recombination 
with electrons. In that case, the neutral-to-protonated abundance ratio at steady state is simply given by the ratio 
of rate constants of the reaction of dissociative recombination and of proton transfer multiplied by the abundance 
ratio between electrons and the proton donor. This simple scheme should hold unless there are important alternative 
ion-molecule reactions of formation of the ion other than proton transfer to the neutral.

For HC$_3$S$^+$ the chemical model indicates that the main formation reactions are proton transfer to C$_3$S from 
HCO$^+$ and H$_3$O$^+$, although the reactions S$^+$ + C$_3$H$_2$ (cyclic or linear) and S + C$_3$H$_3^+$ 
(cyclic or linear) are also efficient at the same level. 
The reactions of S$^+$ with both cyclic and linear C$_3$H$_2$ have been calculated to be 
barrierless when leading to HC$_3$S$^+$ \citep{Redondo1999}. However, the cyclic 
isomer of C$_3$H$_2$ is 28 times more abundant than the linear one in TMC-1 \citep{Fosse2001}, 
and thus the reaction S$^+$ + c-C$_3$H$_2$ will contribute significantly more than S$^+$ + H$_2$C$_3$ 
to the formation of HC$_3$S$^+$.
The fact that the 
calculated C$_3$S/HC$_3$S$^+$ ratio is in agreement with the observed value supports the hypothesis that HC$_3$S$^+$ is 
formed by the aforementioned reactions.

In the case of HCS$^+$, the chemical model underestimates its formation, which is carried out by proton transfer 
to CS from H$_3$O$^+$ and HCO$^+$ and by the reaction between CS$^+$ and H$_2$. Including grain-surface chemical 
reactions, \cite{Vidal2017} calculate a CS/HCS$^+$ ratio in better agreement with observations, especially at late 
times. The reason for the low abundance of HCS$^+$ in our gas-phase model may be related to the underproduction of 
H$_2$S (see e.g. \citealt{Agundez2013}), which is enhanced when including grain-surface chemistry, and which can 
increase the abundance of HCS$^+$ through reactions like H$_2$S + C$^+$ and H$_2$S$^+$ + C.

While the calculated abundance of HC$_2$S$^+$ is in agreement with the observed upper limit, the calculated 
C$_2$S/HC$_2$S$^+$ ratio is too low because the chemical model
severely underestimates the abundance of C$_2$S. 
This is because the chemical network {\small \texttt{kida.uva.2014}} includes an efficient destruction channel 
for C$_2$S through reaction with O atoms, based on estimations by \cite{Loison2012}. If this reaction is neglected, 
the calculated abundance of C$_2$S is shifted up and becomes very close to the observed value, while the abundance 
of HC$_2$S$^+$  also increases, remaining consistent with the observed upper limit (see dotted lines in 
Fig.~\ref{fig:model}). Accurate calculations of the rate constant of the O + C$_2$S reaction at low temperatures 
are needed to shed light on the chemistry of C$_2$S in cold dense clouds.

\section{Conclusions}

We report the first identification in space of protonated C$_3$S. Four harmonically related lines observed 
toward TMC-1 using the Yebes 40m radiotelescope have been unambiguously assigned to this ion thanks to accurate 
\textit{ab initio} quantum chemical calculations and laboratory measurements of the rotational spectrum of this species. 
The derived C$_3$S/HC$_3$S$^+$ ratio of 55 $\pm$ 20 is well reproduced by a gas-phase chemical model in which HC$_3$S$^+$ 
is mostly formed through protonation of C$_3$S and the reactions S$^+$ + C$_3$H$_2$ and S + C$_3$H$_3^+$. 

\begin{acknowledgements}

We thank Ministerio de Ciencia e Innovaci\'on of Spain (MICIU) for funding support through projects 
AYA2016-75066-C2-1-P, PID2019-106110GB-I00, PID2019-107115GB-C21 / AEI / 10.13039/501100011033, and
PID2019-106235GB-I00.
We also thank ERC for funding 
through grant ERC-2013-Syg-610256-NANOCOSMOS. M.A. thanks MICIU for grant RyC-2014-16277. Y.E. thanks 
Ministry of Science and Technology of Taiwan through grant MOST108-2113-M-009-25.

\end{acknowledgements}

\begin{appendix}
\section{Column densities under LTE for linear molecules}
\label{LTE}
The determination of column densities from the observed line paramaters
of the rotational transitions of a molecule requires knowledge of
the collisional rates of the molecular species under study. These values
are not always available and rotational diagrams are used to derived
rotational temperatures, $T_r$, and column densities \citep{Goldsmith1999}.
Often only one line has been observed and therefore an assumption has to
be made as to the rotational temperature. The line parameters of the species
studied in this paper are given in Table \ref{tab_s_bearing}.

In this section we analyse how 
the derived column density depends on the adopted $T_r$ for a range of
frequencies and upper energy levels. 
Let us assume an optically thin line ($\tau\ll1$) with frequency $\nu_{ul}$ and
arising from an upper level with energy $E_u$. If the medium has uniform excitation conditions, 
and the molecular levels are populated  under a uniform rotational temperature $T_r$, then,
the observed brightness temperature ($T_B$) is given by:
\begin{equation}
T_B(v)=(T_r-T_{bg}) \tau(v)
,\end{equation}
where T$_{bg}$ is the temperature of the cosmic radiation background, and
$\tau(v)$ the opacity of the line at frequency $\nu$, which 
is given by
\begin{equation}
\tau(\nu)= \frac{c^2A_{ul}g_u}{8\pi\nu^2}\left(\frac{N_l}{g_l}-\frac{N_u}{g_u}\right) \phi(\nu)
,\end{equation}
\noindent
where $N_l$, $N_u$, $g_l$, and $g_u$ are the column densities and statistical weights of the lower and upper 
levels, $A_{ul}$ is the Einstein coefficient of the transition, and $\phi(\nu)$ is the normalised
line profile, which for a Gaussian line is given by
\begin{equation}
\phi(\nu)=\frac{c}{\nu_{ul}\sigma\sqrt{\pi}} e^{-(\frac{\nu-\nu_{ul}}{\sigma} \frac{c}{\nu_{ul}})^2}
,\end{equation}

\noindent
where $\sigma$=$\Delta v$/(2$\sqrt{ln2}$), with $\Delta\nu$ being the line full width at half intensity
(in km s$^{-1}$). Using the 
following relations in which $S_{ul}$ is the line strength and $\mu$ is the permanent dipole moment of the molecule,
\begin{equation}
A_{ul}=\frac{64\pi^4\nu_{ul}^3}{3hc^3}\frac{S_{ul}\mu^2}{g_u}
,\end{equation}
\begin{equation}
\frac{N_u/g_u}{N_l/g_l}=e^{-h\nu_{ul}/k_BT_r}
,\end{equation}
\begin{equation}
N_u=\frac{N}{Q_{rot}}g_ue^{-E_u/k_BT_r}
,\end{equation}
\noindent
we derive
\begin{equation}
\tau(\nu=\nu_{ul})= \left(\frac{8\pi^{5/2}S_{ul}\mu^2}{3h \sigma}\right)\left(\frac{N}{Q_{rot}}\right) e^{-E_u/k_BT_r} (e^{h\nu_{ul}/k_BT_r}-1) 
,\end{equation}

\noindent
where $N$ is the column density of the molecule, $Q_{rot}$ is the rotational partition function, $h$ is
the Planck constant, and $k_B$ is the Boltzmann constant. 
For linear molecules, the rotational partition function \citep{Gordy1984} is
given by:
\begin{equation}
Q_{rot}\simeq\frac{k_BT_r}{hB}
,\end{equation}
where $B$ is the rotational constant of the molecule. This approximation is rather accurate, even
for low values of $T_r$ if $T_r/B\gg$1. Hence, the brightness temperature at the line centre ($\nu=\nu_{ul}$) is given by 
\begin{equation}
T_B(\nu_{ul})\simeq\frac{(T_r-T_{bg})}{T_r}\left(\frac{8\pi^{5/2}S_{ul}\mu^2B}{3\sigma k_B}\right) N  e^{-E_u/k_BT_r} (e^{h\nu_{ul}/k_BT_r}-1)
,\end{equation}
and the column density can be derived from the observed brightness temperature of the observed transition
from the following expression:
\begin{equation}
N\simeq C\;T_B(\nu_{ul})\;\Delta v\; f(E_u,\nu_{ul},T_r)
,\end{equation}
where $C$ is a constant that depends on the molecular parameters and is given by
\begin{equation}
C=\frac{3k_B}{8\pi^{5/2}S_{ul}\mu^2B\sqrt{4 ln 2}}
,\end{equation}
\begin{equation*}
\;\;\;=\frac{1.7774\times10^{14}}{S_{ul}\mu^2B}\;\;cm^{-2}K^{-1}(kms^{-1})^{-1};
\end{equation*}
\noindent
in these expressions $\Delta v$ is in km\,s$^{-1}$, and corresponds to the line full width at half power intensity
(see expression A.3),
$B$ is in GHz and $\mu$ in D. 

The function $f$ depends on $E_u$, $\nu_{ul}$, and $T_r$ as
\begin{equation}
f(E_u,\nu_{ul},T_r)=\frac{T_r}{(T_r-T_{bg})}\frac{e^{E_u/k_BT_r}}{(e^{h\nu_{ul}/k_BT_r}-1)}
,\end{equation}

\begin{figure}[]
\label{fig_trot}
\includegraphics[width=0.95\columnwidth,angle=0]{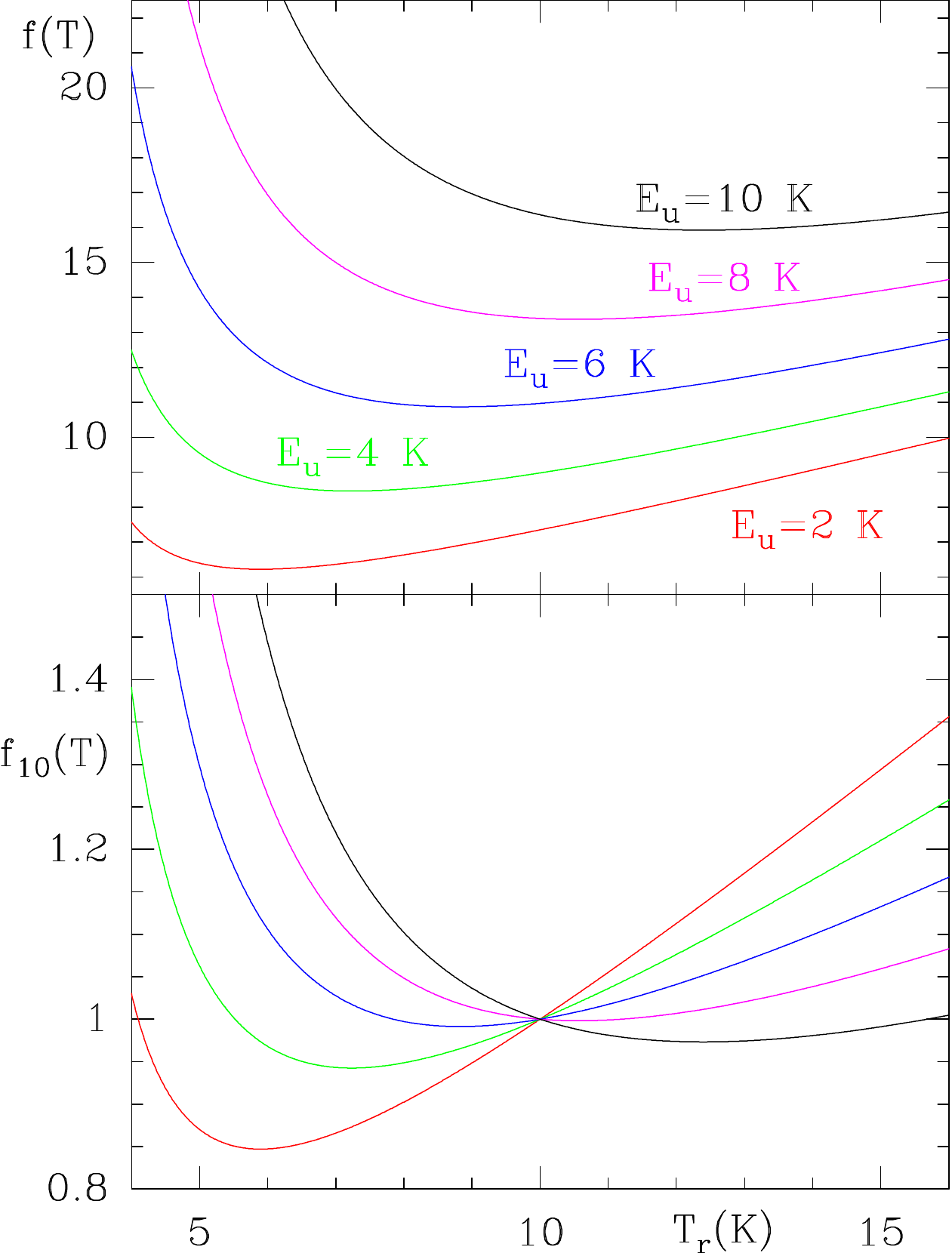} 
\caption{Dependency of the derived column density on the assumed rotational
temperature for a line at 42.674 GHz (HCS$^+$ $J$=1-0; upper panel). The upper energy level is varied between
2 and 10 K. The lower panel shows the same function but normalised to its value for T$_r$=10 K.}
\end{figure}

\noindent
and it shows a smooth behaviour with $T_r$ for most cases of interest in the frequency
range of our survey. Figure A.1 shows the value of the function $f$ for different values of the energy
of the upper level and for a frequency of 42.674 GHz (the frequency of the $J$=1-0 transition of
HCS$^+$). The case $E_u$=2\,K corresponds, specifically, to the $J$=1-0 transition of HCS$^+$. For this
particular case the value of $f$ for $T_r$=5 K is 6.59 while for $T_r$=10 K it is 7.56. Hence, assuming
$T_r$=10 K represents a change in the column density of +15\% with respect the case with $T_r$=5 K. The bottom panel
of Fig. A.1 shows the function $f$ normalised to its value at $T_r$=10 K. Hence, the different plots, which
correspond to different energies of the upper level (always for a frequency of 42.674 GHz), show
the relative error on the estimated column density assuming $T_r$=10 K with respect other values
of the rotational temperature. 

\begin{table*}
\small
\caption{Observed line parameters in TMC-1 for CS, HCS$^+$, C$_2$S and C$_3$S}
\label{tab_s_bearing}
\centering
\begin{tabular}{{lclcccc}}
\hline
Molecule & Transition & \multicolumn{1}{c}{$\nu_{rest}$~$^a$} & $\int T_A^* dv$~$^b$ & $V_{\rm LSR}$~$^c$& $\Delta v$~$^d$ & $T_A^*$~$^e$ \\
         &                      &   \multicolumn{1}{c}{(MHz)}        &  (K\,km\,s$^{-1}$)   & (km\,s$^{-1}$) & (km\,s$^{-1}$) & (K) \\
\hline
CS            & 1-0 & 48990.9573(1)& 1.0623(10)& 5.75(1)& 0.74(1)& 1.3451(14)\\
              &     &               & 0.1031(20)& 6.46(1)& 0.72(1)& 0.1345(14)\\
C$^{34}$S     & 1-0 & 48206.9417(1)& 0.4025(10)& 5.82(1)& 0.65(1)& 0.5902(12)\\
$^{13}$C$^{34}$S&1-0& 45463.4123(6)& 0.0044(10)& 5.72(5)& 0.68(9)& 0.0061(10)\\
HCS$^+$       & 1-0 & 42674.1954(14)& 0.2395(10)& 5.79(1)& 0.54(1)& 0.4154(7)\\
HC$^{34}$S$^+$& 1-0 & 41983.0630(16)& 0.0138(10)& 5.79(1)& 0.58(2)& 0.0223(6)\\
C$_2$S & 2$_3$-1$_2$& 33751.3699(6)& 1.1417(10)& 5.77(1)& 0.75(1)& 1.4346(4)\\
C$_2$S & 3$_3$-3$_2$& 38866.4200(29)& 0.1918(10)& 5.77(1)& 0.63(1)& 0.2842(5)\\
C$_2$S & 4$_3$-3$_2$& 43981.0190(36)& 0.1745(10)& 5.78(1)& 0.59(1)& 0.2780(7)\\
C$_2$S & 3$_4$-2$_3$& 45379.0275(13)& 1.0521(10)& 5.79(1)& 0.60(1)& 1.6468(7)\\
C$_2$$^{34}$S & 2$_3$-1$_2$& 33111.8420(13)& 0.0842(10)& 5.79(1)& 0.72(1)& 0.1101(4)\\
C$_2$$^{34}$S & 3$_3$-2$_2$& 38015.2321(23)& 0.0070(10)& 5.79(1)& 0.63(1)& 0.0100(5)\\
C$_2$$^{34}$S & 4$_3$-3$_2$& 42918.1822(13)& 0.0062(10)& 5.81(1)& 0.55(1)& 0.0108(6)\\
C$_2$$^{34}$S & 3$_4$-2$_3$& 44497.6025(42)& 0.0820(10)& 5.82(1)& 0.57(1)& 0.1343(8)\\ 
C$_3$S        & 6-5 & 34684.3681(6)& 0.4938(10)& 5.80(1)& 0.72(1)& 0.6482(4)\\
C$_3$S        & 7-6 & 40465.0144(7)& 0.4599(10)& 5.81(1)& 0.62(1)& 0.6955(5)\\
C$_3$S        & 8-7 & 46245.6231(8)& 0.4073(10)& 5.82(1)& 0.60(1)& 0.6437(9)\\
C$_3$$^{34}$S & 6-5 & 33844.2468(9)& 0.0198(10)& 5.84(1)& 0.73(1)& 0.0255(4)\\
C$_3$$^{34}$S & 7-6 & 39484.8768(11)& 0.0200(10)& 5.85(1)& 0.64(1)& 0.0296(5)\\
C$_3$$^{34}$S & 8-7 & 45125.4708(12)& 0.0185(10)& 5.84(1)& 0.59(2)& 0.0291(9)\\
\hline
\end{tabular}
\tablefoot{\tablefoottext{a}{Predicted frequencies from the MADEX code \citep{Cernicharo2012}. Numbers in 
parentheses are 1$\sigma$ uncertainties in units of the last digits.}\\ 
\tablefoottext{b}{Integrated line intensity 
in K km\,s$^{-1}$.}\\ 
\tablefoottext{c}{Line velocity with respect to the Local Standard of Rest in km\,s$^{-1}$.} \\
\tablefoottext{d}{Linewidth at half intensity derived by fitting a Gaussian function to the observed 
line profile (in km\,s$^{-1}$).} \\
\tablefoottext{e}{Antenna temperature in K.} \\
}
\end{table*}

For most molecules detected in the survey we observe transitions
with low $J_u$ and therefore with moderate upper energy levels. The error
on the derived column density from the observed line parameters of just one transition of a linear molecule
is below 20\% for $E_u\leq$6 K and for $T_r$ between 5 and 10 K. 
This applies particularly to CS and HCS$^+$ and other linear species in our survey. For transitions
involving levels with $E_u \gg T_r$ the error on the estimated column density
can be considerably larger. From Fig. A.1 we can see that a change from $T_r$= 6 to 10 K introduces
an error of $\sim$1.5 for a transition with $E_u$=10 K. The error can reach a factor three if
$E_u$=15 K. For these cases a realistic estimation of the rotational temperature is needed
based on the observation of several rotational transitions of the molecule under study. 

An additional source of uncertainty in the column density when only one line is observed is 
the assumption of
local thermodynamical equilibrium (LTE) under a uniform rotational temperature for all rotational
levels. 
For the physical conditions of TMC-1, T$_K$=10 K and
n(H$_2$)=4$\times$10$^4$ cm$^{-3}$ \citep{Cernicharo1987,Fosse2001}, the $J$=1-0 line of most molecules 
will have an excitation temperature
close to the kinetic one. However, the excitation of rotational levels with higher values of $J$ will be 
certainly below the kinetic temperature. To quantify this effect we used the Large Velocity Gradient (LVG) 
approximation for CS. We assumed a line width of 0.6 km $^{-1}$ and a column density of 10$^{12}$ cm$^{-2}$ (optically
thin case). We adopted the collisional rates CS/p-H$_2$ provided by \citet{Denis2018}. We obtain T$_{ex}$($J$=1-0)=7.3 K
and T$_{ex}$($J$=2-1)=3.9 K. If the collisional rates CS/o-H$_2$ from the same authors are adopted, then the
derived excitation temperatures for these transitions are 8.9 and 4.9 K, respectively. From
the observed line parameters of the $J$=1-0 transition of C$^{34}$S and $^{13}$C$^{34}$S given 
in Table \ref{tab_s_bearing}, 
we derive
N(C$^{34}$S)=6.5$\times$10$^{12}$ cm$^{-2}$, and N($^{13}$C$^{34}$S)=5.5$\times$10$^{10}$ cm$^{-2}$. These
values are a factor 2.2 and 2.6 lower than those provided in Table \ref{column_densities}, respectively, 
which have been obtained 
assuming a uniform rotational temperature of 10 K. Hence, the main source of uncertainty in the derived
column densities could be related to the assumption of a uniform rotational temperature for all rotational levels
of a linear molecule (LTE conditions), rather than to the adopted rotational temperature.

\section{Observed line parameters and frequency predictions up to $J$=30}
The observed line parameters
for the sulfur-bearing molecules discussed in this paper are given in
Table \ref{tab_s_bearing}. For HC$_3$S$^+$ they are given in Table~\ref{tab_hc3s+}.
Frequency predictions for HC$_3$S$^+$ up to $J$=30 are given in Table~\ref{pred_hc3s+}

\begin{table}
\caption{Predicted line frequencies for HC$_3$S$^+$}
\label{pred_hc3s+}
\centering
\begin{tabular}{cccccc}
\hline
 $J_u-J_l$&     $\nu$(MHz)$^a$      & E$_u$(K)&$A_{ul}$(s$^{-1}$) & $S_{ij}$& $g_u$\\
 \hline
  1- 0    &     5470.9255$\pm$0.0004&  0.3    & 1.901$\times$10$^{-9}$&  1& 3 \\
  2- 1    &    10941.8469$\pm$0.0008&  0.8    & 1.825$\times$10$^{-8}$&  2& 5 \\
  3- 2    &    16412.7601$\pm$0.0011&  1.6    & 6.601$\times$10$^{-8}$&  3& 7 \\
  4- 3    &    21883.6608$\pm$0.0013&  2.6    & 1.623$\times$10$^{-7}$&  4& 9 \\
  5- 4    &    27354.5451$\pm$0.0014&  3.9    & 3.241$\times$10$^{-7}$&  5& 11\\
  6- 5    &    32825.4087$\pm$0.0016&  5.5    & 5.687$\times$10$^{-7}$&  6& 13\\
  7- 6    &    38296.2475$\pm$0.0024&  7.4    & 9.131$\times$10$^{-7}$&  7& 15\\
  8- 7    &    43767.0575$\pm$0.0036&  9.5    & 1.374$\times$10$^{-6}$&  8& 17\\
  9- 8    &    49237.8344$\pm$0.0057& 11.8    & 1.970$\times$10$^{-6}$&  9& 19\\
 10- 9    &    54708.5742$\pm$0.0084& 14.4    & 2.716$\times$10$^{-6}$& 10& 21\\
 11-10    &    60179.2727$\pm$0.0118& 17.3    & 3.631$\times$10$^{-6}$& 11& 23\\
 12-11    &    65649.9258$\pm$0.0160& 20.5    & 4.731$\times$10$^{-6}$& 12& 25\\
 13-12    &    71120.5294$\pm$0.0211& 23.9    & 6.034$\times$10$^{-6}$& 13& 27\\
 14-13    &    76591.0793$\pm$0.0270& 27.6    & 7.556$\times$10$^{-6}$& 14& 29\\
 15-14    &    82061.5714$\pm$0.0340& 31.5    & 9.315$\times$10$^{-6}$& 15& 31\\
 16-15    &    87532.0017$\pm$0.0420& 35.7    & 1.133$\times$10$^{-5}$& 16& 33\\
 17-16    &    93002.3659$\pm$0.0512& 40.2    & 1.361$\times$10$^{-5}$& 17& 35\\
 18-17    &    98472.6599$\pm$0.0615& 44.9    & 1.618$\times$10$^{-5}$& 18& 37\\
 19-18    &   103942.8796$\pm$0.0731& 49.9    & 1.906$\times$10$^{-5}$& 19& 39\\
 20-19    &   109413.0209$\pm$0.0860& 55.1    & 2.226$\times$10$^{-5}$& 20& 41\\
 21-20    &   114883.0796$\pm$0.1003& 60.7    & 2.580$\times$10$^{-5}$& 21& 43\\
 22-21    &   120353.0517$\pm$0.1161& 66.4    & 2.969$\times$10$^{-5}$& 22& 45\\
 23-22    &   125822.9330$\pm$0.1335& 72.5    & 3.396$\times$10$^{-5}$& 23& 47\\
 24-23    &   131292.7193$\pm$0.1524& 78.8    & 3.862$\times$10$^{-5}$& 24& 49\\
 25-24    &   136762.4066$\pm$0.1730& 85.3    & 4.368$\times$10$^{-5}$& 25& 51\\
 26-25    &   142231.9907$\pm$0.1954& 92.2    & 4.917$\times$10$^{-5}$& 26& 53\\
 27-26    &   147701.4674$\pm$0.2196& 99.2    & 5.510$\times$10$^{-5}$& 27& 55\\
 28-27    &   153170.8328$\pm$0.2457&106.6    & 6.149$\times$10$^{-5}$& 28& 57\\
 29-28    &   158640.0825$\pm$0.2737&114.2    & 6.836$\times$10$^{-5}$& 29& 59\\
 30-29    &   164109.2126$\pm$0.3038&122.1    & 7.572$\times$10$^{-5}$& 30& 61\\
 \hline
\end{tabular}
\end{table}
\end{appendix}

\end{document}